\documentclass[11pt,a4paper]{article}

\usepackage[utf8]{inputenc}
\usepackage[T1]{fontenc}
\usepackage{lmodern}
\usepackage[margin=25mm]{geometry}
\usepackage{amsmath}
\usepackage{amssymb}
\usepackage{graphicx}
\usepackage{booktabs}
\usepackage{xcolor}
\usepackage{tikz}
\usetikzlibrary{arrows.meta, positioning, fit, backgrounds, calc}
\usepackage[hidelinks]{hyperref}
\usepackage{authblk}

\newif\iffull
\fulltrue

\newif\ifdraft
\draftfalse
\newcounter{pendientes}
\ifdraft
  \newcommand{\todo}[1]{\stepcounter{pendientes}\textcolor{red}{\textbf{[TODO: #1]}}}
  \newcommand{\reuse}[1]{\stepcounter{pendientes}\textcolor{blue!60!black}{\textbf{[REUSE: #1]}}}
\else
  \newcommand{\todo}[1]{\stepcounter{pendientes}}
  \newcommand{\reuse}[1]{\stepcounter{pendientes}}
\fi

\hypersetup{
  pdftitle={Real-Time Requirements and Transferability in Compton Imaging:
            From the Detector Chain to the Application},
  pdfauthor={F. J. Albiol, E. Larrea Estrelles, S. Tortajada, L. Caballero,
             J. Escalante, A. Albiol, J. L. Leganes-Nieto},
  pdfkeywords={Compton camera; gamma-ray imaging; real-time systems; bounded
               latency; data acquisition; image reconstruction; technology
               transfer; intended use}}

\title{\bfseries Real-Time Requirements and Transferability in Compton
Imaging:\\ From the Detector Chain to the Application}

\author[1]{Francisco~J.~Albiol\thanks{kiko@ific.uv.es}}
\author[1]{Elena~Larrea~Estrelles}
\author[1]{Salvador~Tortajada}
\author[1]{Luis~Caballero}
\author[1]{Jos\'e~Escalante}
\author[2]{Alberto~Albiol}
\author[3]{Jos\'e~Luis~Legan\'es-Nieto}
\affil[1]{Instituto de F\'isica Corpuscular (IFIC), CSIC--Universitat de
Val\`encia, E-46980 Paterna, Val\`encia, Spain}
\affil[2]{Universitat Polit\`ecnica de Val\`encia, E-46022 Val\`encia, Spain}
\affil[3]{ENRESA, Spain}
\date{}

\begin{document}
\maketitle

\begin{abstract}
\noindent
Compton cameras are proposed for tasks whose value decays with delay:
verifying a range during irradiation, guiding an intervention, characterising
an inaccessible volume, surveying a band no telescope covers. Whether a device
can serve such a task is settled by the composition of its whole chain, while
the literature that would answer the question is organised by stage --- so
claims made at application level routinely rest on evidence obtained at
component level.

This review walks that chain, asking at each stage what binds first and what
a reader can determine from the published text, and then asks what governs
whether a capability transfers between groups and between application domains.
The evidence is of two kinds: 83 full texts scored in context against defined
markers, and, for five application domains measured alike, the size of the
receiving literature, of the need it states, of the incumbent and of the work
in service.

Performance is reported at one operating point in 65 of 83 full texts, with
count rate swept in none: the field reports values where transfer requires
gradients. An accelerator is used in 39 works and a learned model in 31, while
separability is discussed in 28, a memory footprint given in 9, and the cost of
a precomputation or the inference time of a model in none. Across domains,
neither the size of the receiving literature, nor of the incumbent, nor of the
stated need orders the domains as deployment commitment does, while whether the incumbent
can serve the task at all, and how many domain boundaries the output must
cross, do so consistently. The distribution of publication is close to the inverse of the
distribution of deployment commitment.

We give the quantities a report must contain for a third party to judge
whether a method fits an application it was not built for.
\end{abstract}


\section{Introduction}
% ===========================================================================
Gamma-ray imaging is proposed for tasks whose value decays with delay. A
range verified while the beam is still on can change the beam; verified
afterwards it can only inform the next fraction. A hotspot located during a
survey redirects the survey; located later it enters a report. Whether an
instrument can serve a task of that kind is settled by the composition of its
whole chain --- detection, digitisation, calibration, event building,
reconstruction, and the delivery of a result to whoever acts on it --- while
the literature that would answer the question is organised by stage, each with
its own units, its own baselines and its own idea of what counts as an
improvement.

This review asks, at each stage, two things: what binds first, and what a
reader can determine from the published text. The first is physics and
engineering. The second is a property of reporting, and it is the one that
decides whether a result can be carried anywhere.

Two kinds of review already exist here, and this one sits between them. The
first addresses the internal development of the method: Kim and
Lee~\cite{kim2024review} set out the principles of Compton imaging and the
reconstruction methods built on them, from analytical inversion through
iterative and statistical estimation to learned approaches, and a reader
entering that stage should begin there. The second addresses the range of
measurement techniques proposed for a particular application, comparing what
each can deliver: for in-vivo range verification, Parodi and
Polf~\cite{parodi2018invivo} survey positron emission, prompt gamma, magnetic
resonance and ionoacoustics side by side. Both are thorough within their frame,
and \emph{this review competes with neither: it takes both as boundaries}.

The question between them is the one addressed here: \textbf{what evidence is
required to determine whether a technical capability can satisfy an imaging
problem and be transferred into use.} A method review establishes what a
technique can do; an application review establishes which techniques are
candidates for a task. Neither establishes what a reader must be told in order
to decide, for their own rate, source and window, whether a reported result
carries. Clinical adoption is one demanding instance of that question rather
than its defining scope, which is why the domains examined here include site
characterisation and astrophysics alongside particle therapy.

The distinction shows in what each kind of survey does for a reader. A survey
of techniques is enumerative: it is complete when every method appears, it
grows as methods are added, and a reader who finishes it knows what exists. A
survey of limits and conditions of use selects: it states, for each approach,
the operating point at which it was demonstrated, the resource that binds
there, and the range over which the reported figure holds --- so a reader who
brings their own conditions finds part of the catalogue excluded and the rest
ordered. The first answers what has been done. The second is what a designer
needs in order to commit a crystal, a channel count and a readout before
anything can be measured.

\iffull
\subsection{How a Compton Camera Forms an Image}
\label{sec:foundational}

Gamma radiation cannot be focused: no lens or mirror bends it, so the origin of
a photon must be inferred from how it interacts rather than from where it is
brought to a point. A collimated camera solves this mechanically, admitting
only photons whose direction agrees with a bored channel and discarding the
rest. A Compton camera replaces that mechanism with a measurement.

The event is a single scatter followed by an absorption
(Fig.~\ref{fig:cone-schema}). A photon of energy $E_0$ scatters in a first
detector, depositing $E_1$ and continuing along a new direction; it is absorbed
in a second detector, depositing $E_2$, with $E_0 = E_1 + E_2$ when the deposit
is complete. The Compton relation fixes the scattering angle from the two
energies alone,
\begin{equation}
 \cos\omega \;=\; 1 - m_{e}c^{2}\left(\frac{1}{E_{2}} -
 \frac{1}{E_{1}+E_{2}}\right),
\label{eq:compton-angle}
\end{equation}
with $m_{e}c^{2} = 511$~keV. The two interaction positions fix an axis. Angle
and axis together define a cone whose apex is the first interaction and whose
surface contains every direction from which the photon could have arrived.

A single event therefore constrains the source to a surface rather than to a
point, and this is the property that governs everything downstream. The source
is recovered where many such surfaces agree, so the information lives in the
intersection of a population rather than in any member of it. Two consequences
follow immediately and are used throughout: the cone carries \emph{two
independent determinations} --- its axis from geometry, its opening angle from
energy --- which degrade for different reasons and are improved by different
expenditure; and every event must be propagated as a surface against a volume,
which sets a lower bound on the cost of any inversion.

The standard figure of merit for a single event is the angular resolution
measure, the difference between the reconstructed cone angle and the true
angle subtended by a known source at the interaction point. It aggregates
energy resolution, position resolution and the irreducible Doppler broadening
of the bound electron's momentum into one number, which is convenient and, as
Section~\ref{sec:building} shows, is also where two separately actionable terms
become one that is not.

\begin{figure}[t]
\centering
\includegraphics[width=\linewidth]{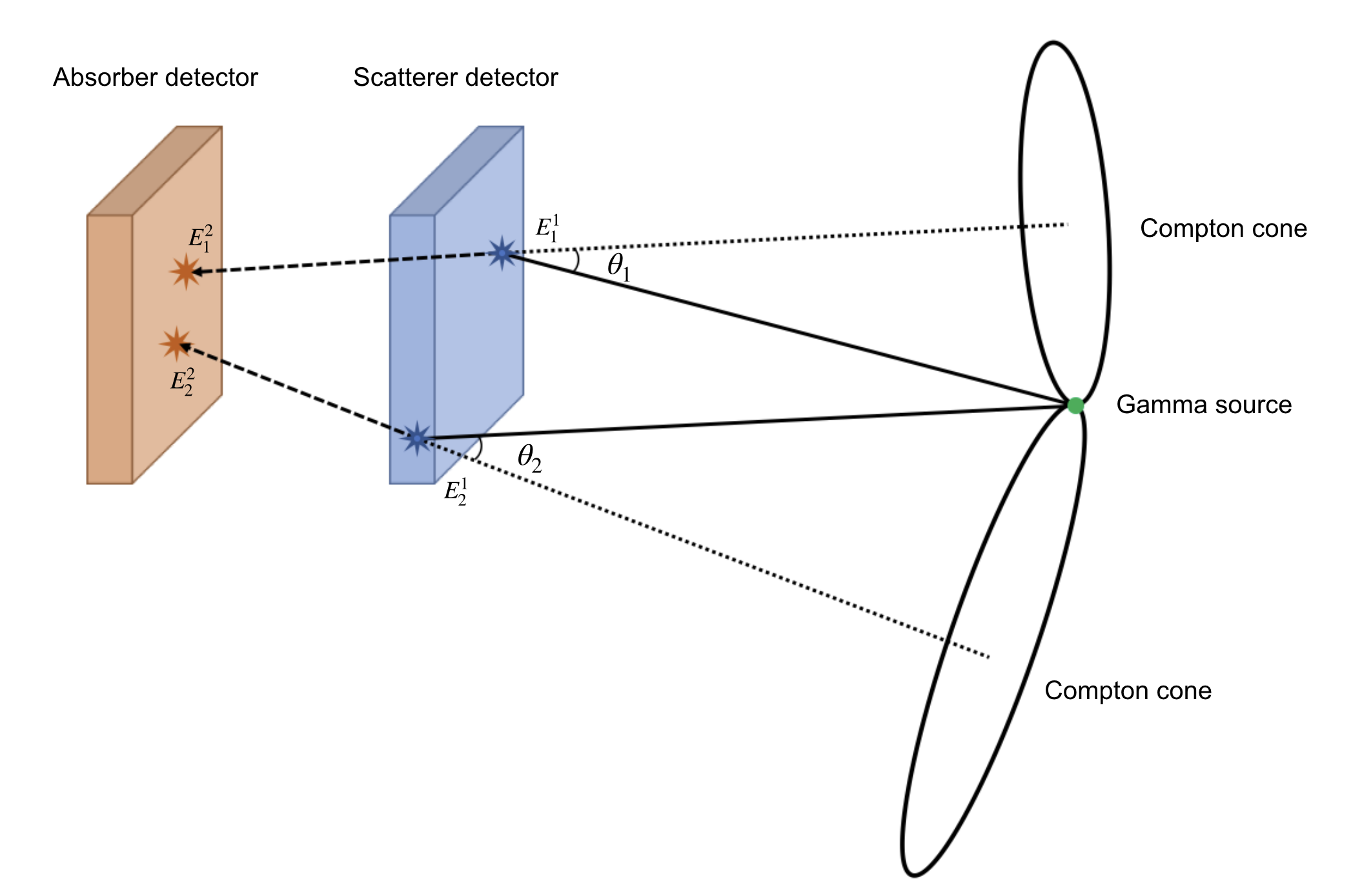}
\caption{Two Compton events from the same source. Each photon scatters in the
first detector and is absorbed in the second; $E^{\,j}_{\,i}$ denotes the
energy deposited by event $i$ in detector $j$. For each event the two
interaction positions fix the cone axis and the two energies fix the opening
angle $\theta_i$ through Eq.~\ref{eq:compton-angle}, so a single event
constrains the source to a conical surface rather than to a point. The source
is recovered where surfaces from different events agree, which is why the
information lives in the intersection of a population rather than in any member
of it. Reproduced from~\cite{albiol2026bounded}.}
\label{fig:cone-schema}
\end{figure}
\fi

Conditions are what transfer: a quantity that holds only inside the stage that
produced it can be enumerated, while a quantity bound to an operating point can
be carried.

% ===========================================================================
\section{What an Imaging Problem Requires}
\label{sec:rank}
% ===========================================================================
An image is the means by which a problem is resolved: where the beam stopped,
whether the margin is clear, which container to open, where not to send a
person. The quantities that decide whether a device can serve such a problem
belong to the problem before they belong to the instrument, and they have to be
available in design --- crystal, channel count, geometry and readout are
committed long before anything can be measured, so a purpose that becomes
quantifiable only in retrospect can be discovered but cannot be specified to a
manufacturer. The consequence for a literature is that a field can report
improvements indefinitely, each real, and remain unable to say whether any
combination of them yields an instrument that serves a stated task: the
improvements are quoted against previous work while the task is stated in other
units. The budget assembled here exists to make that accumulation composable,
every term being a quantity that can be fixed on paper beforehand and held to
afterwards.

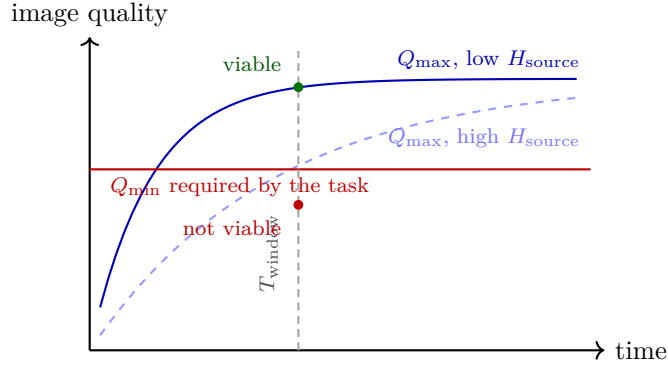
\begin{figure}[t]
\centering
\begin{tikzpicture}[scale=0.92]
\draw[->,thick] (0,0) -- (7.4,0) node[right,font=\small] {time};
\draw[->,thick] (0,0) -- (0,4.5) node[above,font=\small] {image quality};
% Qmax curves for two source complexities
\draw[thick,blue!70!black] plot[domain=0.15:7,samples=60]
      (\x, {3.9*(1-exp(-1.15*\x))});
\node[blue!70!black,font=\scriptsize,anchor=east] at (7.2,4.18)
      {$Q_{\max}$, low $H_{\mathrm{source}}$};
\draw[thick,blue!40,dashed] plot[domain=0.15:7,samples=60]
      (\x, {3.9*(1-exp(-0.38*\x))});
\node[blue!55,font=\scriptsize,anchor=east] at (7.2,3.05)
      {$Q_{\max}$, high $H_{\mathrm{source}}$};
% Qmin
\draw[thick,red!75!black] (0,2.6) -- (7.2,2.6);
\node[red!75!black,font=\scriptsize,anchor=west] at (0.15,2.35) {$Q_{\min}$ required by the task};
% window
\draw[thick,gray!70,dash pattern=on 3pt off 2pt] (3.0,0) -- (3.0,4.3);
\node[gray!60!black,font=\scriptsize,rotate=90,anchor=south] at (2.85,1.4) {$T_{\mathrm{window}}$};
% viability marks
\fill[green!45!black] (3.0,3.78) circle (2.0pt);
\node[green!35!black,font=\scriptsize,anchor=south east] at (2.90,3.86) {viable};
\fill[red!75!black] (3.0,2.09) circle (2.0pt);
\node[red!70!black,font=\scriptsize,anchor=north east] at (2.90,2.02) {not viable};
\end{tikzpicture}
\caption{The viability bracket. The task fixes $Q_{\min}$; the system and the
source fix $Q_{\max}$ as it grows with accumulated measurement. An application
is viable where the attainable quality clears the required quality inside the
available window, and a rise in source complexity can move the same system from
one side of that test to the other without any change to the algorithm.}
\label{fig:bracket}
\end{figure}

Four questions found the problem, and everything a purchaser, a regulator or a
referee will ask reduces to one or two of them. They are also the sequence in
which a claim is examined --- an intended use, a performance claim, and the
evidence supporting it --- which is what makes the requirement legible to
whoever pays for a device and not only to whoever builds it. The four are: \textbf{Q1}, what decision the
image must support; \textbf{Q2}, the window in which it must be available for
that decision to remain actionable; \textbf{Q3}, the quality below which the
decision cannot be taken; and \textbf{Q4}, the extent and complexity of what is
being imaged. Each is answered by a quantity rather than by an argument, and naming the
quantity is what turns the question into a specification: Q1 is answered by the
consumer of the image and the decision it takes, Q2 by a bound stated per
operating state, Q3 by $Q_{\min}$ expressed in the units the receiving
discipline already uses, and Q4 by the extent and complexity of the source
together with the rate at which performance degrades as they grow.
\iffull   % ---- solo arXiv: tab:problem, la tabla de dimensiones de evaluacion
Table~\ref{tab:problem} sets out the dimensions on which an
imaging problem is actually evaluated, the founding question each belongs to,
and where each one lands among the fronts.

\begin{table}[t]
\centering
\caption{What an imaging problem is evaluated on, which founding question each
dimension belongs to, and where it appears among the fronts. Every dimension
reduces to one or two of the four questions, which is what makes the list
finite.}
\label{tab:problem}
\footnotesize
\begin{tabular}{@{}p{0.335\linewidth}p{0.10\linewidth}p{0.44\linewidth}@{}}
\toprule
\textbf{Dimension of evaluation} & \textbf{Question} & \textbf{Where it lands} \\
\midrule
Time against the incumbent technology & Q1, Q2 &
The window, and the comparison that justifies the device at all \\
\addlinespace
Expected improvement over current practice & Q1, Q3 &
Sets $Q_{\min}$: the improvement has to be one the decision can use \\
\addlinespace
Spatial resolution & Q3, Q4 &
Low- and high-resolution fronts; fixes the cell and thence the statistics \\
\addlinespace
Contrast and signal to noise & Q3, Q4 &
Source-entropy fronts; degrades as the activity spreads \\
\addlinespace
Quantification of activity, such as uptake & Q1, Q3 &
Requires live time and loss accounting: observability in
Table~\ref{tab:comp}, without which a number cannot be corrected \\
\addlinespace
Limits on capture time & Q2 &
Low-rate front: the window and the statistics meet here \\
\addlinespace
Interaction with other equipment: totalisers, displays, alert devices & Q1, Q2 &
The consuming system, which is where the decision is taken and therefore where
the window is set \\
\addlinespace
Time series of repeated measurements & Q2, Q4 &
Bounded state: each measurement must begin from a defined condition rather
than inherit the last \\
\addlinespace
Long-exposure measurement & Q3, Q4 &
Bounded state and calibration drift; the regime where an unbounded term is
discovered rather than declared \\
\bottomrule
\end{tabular}
\end{table}
\else
% BRIDGE (texto nuevo, solo version revista)
Every dimension on which such a problem is evaluated reduces to one or two of
the four, which is what makes the list finite.
\fi

\iffull   % ---- solo arXiv: sec:rank, el encuadre (que capacidad es y a quien no juzga)
The capability that makes this instrument worth the trouble deserves naming
before requirements are imposed on it. Recovering a distribution in three
dimensions from a set of cones is the reconstruction of a scene from its
collisions: each event contributes a surface rather than a point, and the
intersection of enough surfaces recovers a volume no collimator reaches without
discarding the photons that carry it. That capability is real and unusual. What
does not follow from it is a purpose --- which scene, for whom, to what
sufficiency, within what time.

These questions are not a standard against which research should be judged.
Work that does not hold them is science: it establishes what is determinable
and requires no application to justify it. They become binding at one moment
--- when a work claims an application --- because such a claim asks somebody
else to act on the result. A reader who holds them can tell which of two
otherwise similar works offers an instrument and which offers a finding, and
that distinction is what this review is organised to make.
\fi

A chain has \emph{bounded latency} when, for a stated configuration and a
stated operating state, an upper bound on the interval between an interaction
in the detector and the availability of its contribution in a \emph{named}
product can be derived from the configuration and confirmed by measurement.
Naming the product matters, since a live spectrum, a Compton candidate and a
reconstruction snapshot have different bounds. Three distinctions make the definition
non-trivial. \emph{Bounded is not small}: a declared bound of two seconds is
more useful in a critical setting than an undeclared typical value of fifty
milliseconds, because only the first can be reasoned about beforehand.
\emph{A bound is not an average}: a mean describes a distribution, a bound
describes the tail, and it holds only in the operating state claimed --- so
degraded, overload, loss and reconfiguration states are part of the claim.
\emph{A bound is not a property of a kernel}: it belongs to the composed chain
and accumulates,
\begin{equation}
 \begin{aligned}
 L_{\mathrm{total}} \;\le\;& L_{\mathrm{transport}} + L_{\mathrm{order}}
   + L_{\mathrm{group}} + L_{\mathrm{pair}} \\
 &+ L_{\mathrm{queue}} + L_{\mathrm{encode}} + L_{\mathrm{recon}},
 \end{aligned}
\label{eq:latency-budget}
\end{equation}
where every term must be declarable in advance from the configuration and
verifiable afterwards by measurement. A term that cannot be declared is not a
small term but an unbounded one, and it makes $L_{\mathrm{total}}$ unbounded
however fast the remaining stages run.

Latency is not the wait for statistics, and losing that separation is much of
why \emph{real time} is used without content here. Two intervals are being
added: the time until the measurement carries enough information for the image
to form at all, which is physics and which no computing changes, and the time
taken to respond once it does, which is the only one a system controls. Where
the distinction is made it is made in these terms, as the existence of windows
that allow interference while delivery is under
way~\cite{neishabouri2025realtime}. It also states what a streaming architecture
buys, provided two quantities are kept apart. Let $C(N)$ be the cost of
inverting an accumulated batch of $N$ events, which grows with $N$, and let
$L_{\mathrm{total}}$ remain what Eq.~\ref{eq:latency-budget} defines: the bound
on a single contribution traversing the chain. If inversion begins only after
accumulation ends, the two intervals add; if encoding and inversion proceed
concurrently with acquisition, the last event still arrives at
$N/R_{\mathrm{acc}}$ and still has to traverse the chain, so
\begin{equation}
 T_{\mathrm{batch}} \;\approx\; \frac{N}{R_{\mathrm{acc}}} + C(N),
 \qquad
 T_{\mathrm{stream}} \;\approx\; \frac{N}{R_{\mathrm{acc}}}
 + L_{\mathrm{total}},
\label{eq:overlap}
\end{equation}
and the gain is $C(N) - L_{\mathrm{total}}$. Streaming does not save a fixed
amount: it replaces a term that grows with the event count by a residual that
does not, which is why the advantage widens exactly where the statistical wait
is longest.

Equation~\ref{eq:overlap} holds under one condition that must be stated with
it. The overlap requires the chain to consume events at least as fast as they
arrive; where the service rate falls below $R_{\mathrm{acc}}$ a queue grows
without bound and the completion time is governed by the service rate instead,
worse than either term above. Throughput and latency are therefore separate
requirements, and a bound on the second is meaningful only in an operating
state where the first is met --- which is the reason
Section~\ref{sec:rank} asks for the bound to be declared per state rather than
once.

A useful chain emits several products with different bounds and different
evidential weight, and the practical question is which decision each is
authorised to support. A fast coarse product can \emph{inform} that something
departs from expectation; a later one, formed from more statistics, can
\emph{confirm} it; a further one can \emph{correct} a delivery or
\emph{formalise} a record. Three latencies attached to three authorities, whose
conflation is what makes a single reconstruction time uninformative. For a tier
$k$ requiring $N_k$ accepted events at an accepted rate $R_{\mathrm{acc}}$,
\begin{equation}
 T_k \;\ge\; \frac{N_k}{R_{\mathrm{acc}}} \;+\; L_{\mathrm{total}}(k).
\label{eq:decision-time}
\end{equation}
The three terms belong to three disciplines, which is the structural reason
they are seldom stated together: $N_k$ follows from the decision to be
supported and the confidence it requires; $R_{\mathrm{acc}}$ from detector
efficiency, geometry, source activity and the selections applied along the way;
$L_{\mathrm{total}}$ from what a representation preserves and how cost scales
with the discretisation the measurement supports. A study conducted entirely
within one frame can be complete on its own terms and still leave the loop
unevaluable, because the loop is the composition of the three.

Together these close a question normally left open. Viability is decided
neither by latency, nor by image quality, nor by statistics separately, but by
whether two bounds meet: the minimum quality that supports the decision,
$Q_{\min}$, which follows from what the image is for and is a property of the
application; and the maximum attainable inside the available window,
\begin{equation}
 Q_{\max}\bigl(T_{\mathrm{window}},\, H_{\mathrm{source}},\,
 R_{\mathrm{acc}},\, L_{\mathrm{total}}\bigr),
\label{eq:qmax}
\end{equation}
which is a property of the system and of the source it faces. The application
is viable where $Q_{\max} \ge Q_{\min}$ inside $T_{\mathrm{window}}$
(Fig.~\ref{fig:bracket}), however good the algorithm and however fast the
kernel. Stated this way the comparison is computable, and it is almost never
computed --- because the two bounds belong to different disciplines, and
neither discipline is expected to supply the other's.

\subsection{Two Imaging Regimes}
\label{sec:regimes}

The chain is shared by two regimes that behave differently at almost every
stage, and a claim that does not say which one it belongs to cannot be
evaluated. In the \emph{near field} the source lies within reach of the
detector: cones from different interaction points intersect at a finite
distance, depth is recoverable, and reconstruction is volumetric. Proximity is
what makes depth available and also what degrades the projection, since the
closer the source the larger the angular error a given position uncertainty
produces. In the \emph{far field} the cones no longer converge at a finite
depth; what survives is direction, the reconstruction is an angular map, and
depth is recovered, if at all, by moving the camera. The first is the regime of
range verification, intraoperative guidance and close-range package
characterisation; the second of contamination survey and decommissioning.

\iffull   % ---- solo arXiv: sec:regimes, el caso reducido a 2D
A third case is a choice rather than a regime. Several uses reduce the
reconstructed object deliberately --- a fall-off position along the beam, a
planar map for an intraoperative probe, a projection for a thyroid or
sentinel-node study --- where the source is in the near field and depth is in
principle available but is not reconstructed, because the task needs a
coordinate and does not have time. The trade is computable rather than
qualitative: collapsing a volume to a plane divides the resolvable source
elements by the depth extent over the cell size, and with them both the
statistics the decision requires and the cells each event visits. A
two-dimensional answer is a different point on the same surface, reached on
purpose, and should be reported with the decision it supports and the depth
knowingly given up.
\fi

We take the near-field volumetric case as general because it is the most
demanding of the three and the other two are computable reductions of it:
collapse a dimension, or send the source to infinity, and the terms follow.
Evidence does not travel in that direction --- a result established in the
reduced case says little about the general one --- which is the same asymmetry
that makes a point-source figure a poor guide to an extended source. The
distinction reaches every term here: $M_{\mathrm{source}}$ and the statistics it
demands, the discretisation and the cost per event, and the decision the image
supports, which fixes $Q_{\min}$.

\subsection{Where the Premise Currently Fails}
\label{sec:premise}

Deriving a requirement from an application assumes the application has stated
one, and two properties of the present situation qualify that assumption.

\iffull   % ---- solo arXiv: sec:premise, los dos argumentos cualitativos
The values a receiving discipline works to are empirical. They were arrived at
by accumulated practice, and they carry the authority that accumulation confers:
a tolerance in clinical use encodes what has been observed to work across many
treatments, in a setting where the consequences of the alternative were
visible. An instrument proposing to change one of those values argues from a
model against a record, which is a harder argument than a performance
comparison and a different one. It is also the argument that would have to be
made, so making it explicitly is worth more than improving the model quietly.

The uses that would justify this instrument most strongly are the ones not yet
in practice, and their requirements are correspondingly unstated. A more
demanding use is likely to ask for something simpler of the act being monitored
rather than something more elaborate, since the act already carries as much
apparatus as it can hold; what that simpler thing is, and what purpose it
serves, remains to be named. A review can mark the position without filling it,
and marking it is the more honest service: the requirement for a future
application is derivable once someone states the application, and until then a
claim aimed at it is aimed at a target of its own construction.
\fi

There is a measurable difficulty behind both. The premise of this section asks
the instrumentation literature to take its requirements from the discipline
that will use the result, and that traffic is currently thin. Across the 58
works in the corpus whose reference lists could be parsed, citations to
venues in which clinical outcomes are reported number 47 against 971 to
instrumentation venues, and 34 of the 58 cite no such venue at all. The recent
statements of need made by the
receiving community --- what it is currently unable to do, and what it has
lately decided it wants --- are largely absent from the literature that
proposes to serve it. That is a break in the premise this section rests on, and
it is stated here rather than in the discussion because it conditions
everything that follows: a chain designed against an unstated requirement is
being designed against an assumed one.

\subsection{Scope, and the Boundary Conditions It Leaves Out}
\label{sec:scope}

This review covers the path from detection to image formation. The image is a
step towards an end, and several conditions that decide whether that end is
reached sit outside the path and inside the design: where the patient or the
object is placed relative to the camera; the interface through which a result
is presented and read, whether a console or a surgical display; the reporting and control loop that carries a decision
back to the treatment or operation being monitored; and the background a real
environment supplies, whether accelerator-induced activation in a therapy room
or the ambient field of a contaminated space.

\iffull   % ---- solo arXiv: sec:scope, por que cada condicion importa
Each of them fixes a term the review does use, which is why they are named
rather than passed over. Placement fixes the near-field geometry and with it
the projection error and the solid angle, so it enters both $R_{\mathrm{acc}}$
and the achievable quality. The interface and the reporting loop are where the
decision is taken, and therefore what fixes the window and $Q_{\min}$.
Environmental background adds to the accepted rate without adding to the
signal, so it consumes acquisition capacity as the intrinsic activity of a
crystal does, and it shifts the statistics a decision requires.
\fi

The boundary is one of subject matter rather than of importance. A design that
treats these as external will meet them at commissioning, when the quantities
they govern have already been committed.

% ===========================================================================
\section{The Chain, End to End}
\label{sec:chain}
% ===========================================================================

Walking the chain serves two purposes at once. It gives a reader who does not
already know it a map of what each stage does, and it exposes a discrepancy
that organises much of this review: \emph{architectural complexity and
computational cost are different quantities, and the literature is distributed
according to the first}. Acquisition and event building are the clearest case.
They are intricate as architecture --- windows, buffers, coincidence logic,
ordering across boundaries --- and they are stable and cheap as computation,
with a cost per event that is bounded and predictable once the policy is
fixed. A great deal is written about them because they are hard to build, and
comparatively little of what is written is a computational quantity.

For each stage below we give what it does, what it costs in architecture and in
computation, how abundant and how varied the published solutions are, and which
properties of Table~\ref{tab:comp} bind there. Table~\ref{tab:abundance}
summarises the abundance. Descending to implementation at each stage is what
separates a statement of the problem from a treatment of it.

\begin{table}[t]
\centering
\caption{Where the published effort sits. Counts are works in the corpus of
Section~\ref{sec:survey} that address each stage; a work may address several.
Abundance and variety are read together: a stage with many works and few
distinct approaches is settled, one with many of both is open.}
\label{tab:abundance}
\footnotesize
\begin{tabular}{@{}p{0.40\linewidth}rp{0.36\linewidth}@{}}
\toprule
\textbf{Stage} & \textbf{Works} & \textbf{Character} \\
\midrule
Detection and front-end & 76 & Abundant, materially varied, maturing \\
Digitisation and transport & 56 & Abundant, converged on few ASIC families \\
Calibration and stability & 41 & Mentioned widely, specified in 5 \\
Event building and grouping & 34 & Architecturally hard, computationally settled \\
Image formation & 79 & Abundant and the most varied of all \\
Quantification and application & 49 & Invoked often, specified rarely \\
\bottomrule
\end{tabular}
\end{table}

\subsection{From Detection to Calibrated Events}
\label{sec:detection}

The first three stages belong to one domain. Detection, digitisation and
calibration are built by the same people, committed at the same time, and share
the property that their per-event cost is a lookup. They are treated briefly
here for a further reason: the problems they pose are common to single-photon
imaging generally, they are served by commercial suppliers, and they are
therefore neither specific to this instrument nor what limits its
applications. What they contribute to the argument is a set of ceilings.

Four quantities leave the detector and cannot be recovered afterwards.
Efficiency and solid angle fix the accepted rate; timing resolution fixes the
coincidence window and the randoms it admits; energy resolution fixes the
spectral separation available to a selection; and position uncertainty,
propagated through the cone geometry, fixes the smallest cell that carries
information --- so the detector, not the reconstruction, sets the finest grid
the system is entitled to use. The trades that produce them run against each
other: thickness buys stopping power and the self-absorption of internal decays
with it, area buys solid angle and multiplies channels, proximity buys solid
angle and degrades the projection. Optimisation studies exist and are generally
conducted against efficiency and resolution rather than against the acquisition
and computational terms the same choices
fix~\cite{lee2020optimization,corpus22evaluation2023}. \iffull   % ---- solo arXiv: sec:detection, cristales y fotosensores
Semiconductors give
position by segmentation and energy by direct conversion at the cost of channel
count~\cite{corpus12characteriza2024}; scintillators with silicon
photomultipliers give stopping power and timing, and a monolithic crystal buys
continuous coordinates and depth of interaction while placing an estimator,
increasingly a learned one~\cite{corpus51submillimete2021}, inside the
per-event path. One term is specific to the lanthanum halides and belongs in
any rate budget built from them: their intrinsic radioactivity supplies a
source-independent count rate that grows faster than crystal volume, consumes
acquisition capacity, and sets a floor beneath any low-activity measurement ---
while being, as a line of known energy present in every acquisition, available
as a permanent internal calibration reference.
\fi

\iffull   % ---- solo arXiv: sec:detection, amortizacion y escalado de canales
Computationally this is the most settled part of the chain, and it is settled
by amortisation. The work per event is a lookup --- apply a per-channel
calibration, map a channel to a pixel, attach a timestamp --- while what is
expensive was paid in advance, in the calibration campaign that produced the
tables and, for a monolithic crystal, in training the position estimator. The
per-event cost stays flat as rate rises and as the grid is refined, so a stage
of considerable architectural difficulty contributes little to the latency
budget. What grows is the channel count: scaling means multiplying identical
front ends rather than making one faster, so aggregate rate, bandwidth, power
and the number of calibration parameters scale together while per-channel
figures stay fixed by a device designed once. Two limits belong to the readout
rather than the crystal --- a per-channel rate ceiling with its dead
time~\cite{nadig2021evaluation,corpus52tofpet2013,corpus08672005}, and power,
which decides whether a front end can sit where the application needs it and
which the reconstruction literature never treats as a
resource~\cite{nadig2019power} --- and two structural properties follow. Block
size is a latency parameter before it is a transport convenience, since a stage
cannot emit before its block is complete. And backpressure exists across this
boundary and stops at it: a consumer that falls behind can decline to advance,
while no mechanism asks a source to emit fewer photons.
\fi

The stage appears in 56 works and has converged on a few ASIC families, which
should be read as difficulty rather than settled interest: emitting coherent
timestamped data from thousands of channels without gaps is hard enough that
the field uses the few devices that manage it. \iffull   % ---- solo arXiv: sec:detection, el empaquetado del ASIC
Such a device is essential
complexity solved once and packaged, and what arrives with it is the accidental
residue --- register maps, configuration parameters, calibration tables,
firmware revisions --- which each group then maintains. That packaging is what
makes this stage the entry point for a new group rather than a barrier to one.
Variety in a stage is inversely related to the cost of attempting it, which is
why image formation carries six competing families in the same corpus and
digitisation carries a handful of devices.
\fi

\iffull   % ---- solo arXiv: sec:detection, las dos vidas de la calibracion
Calibration completes the group and comes before events exist: what arrives is
digitised values with timestamps, and the map to an energy and a position is
what calibration supplies. It has two lives easily confused --- parameters
\emph{established} out of band, amortised and expensive, and \emph{applied} in
band as a lookup, continuous and cheap --- and its coupling to the next stage
is direct, since energy calibration sets the opening angle of the cone and
position calibration sets the axis. Linearity decides whether the calibration
holds across the range actually used, giving correct angles for one class of
event and wrong angles for another, silently. Stability is the same property in
time, and it carries a design rule that is easy to violate: a verification is
informative only where its acceptance band is narrower than the effect the
receiving party tracks, and a band chosen wider is formally correct and blind
by construction~\cite{vijande2025drift}.
\fi

The reporting across these three stages is thinnest exactly where scaling
hurts. A channel count appears in 15 works, a power figure in 11, a statement
about scalability in 12, an aggregate bandwidth in 5. Calibration is mentioned
in 41, an energy calibration described in 12, a position calibration in 2,
linearity or a saturation correction in 24, a repeatable procedure in 5, drift
and its correction in 2, and a recalibration interval in none. Per-channel
calibration tables are mentioned in 6 works and their maintenance in one. A
design that multiplies channels multiplies that burden exactly, so the quantity
that scales worst with instrument size is the one least available to anyone
estimating what a larger instrument would cost to keep working.

\subsection{Reconstruction: From Signals to a Distribution}
\label{sec:building}

\subsubsection{The cone is two measurements, reported as one.}
Signals are grouped into interactions, interactions are paired across planes,
and each surviving pair becomes the geometrical object reconstruction consumes.
A cone, however, is not one measurement but two. Its \emph{axis} comes from
geometry, the line joining the two interaction positions; its \emph{opening
angle} comes from energy, through the Compton relation between the deposit in
the scatterer and the energy carried away. They are set by different parts of
the instrument, degrade for different reasons, and are improved by different
expenditure --- position resolution and plane separation against energy
resolution and completeness of deposit --- so a system that spends on the wrong
one improves nothing the image can use. The corpus reports the sum and rarely
the parts: an angular resolution appears in 38 works of 83 and the Compton
relation in 40, Doppler broadening in 30, while the contribution of energy
resolution to angular uncertainty is made explicit in 6 and the geometric
contribution of position uncertainty in none we found. The term that cannot be
engineered away receives the most attention; the two that can be are reported
together.

Grouping decides both determinations, which is what makes it more than
bookkeeping: what is summed fixes the energy, and the pattern used fixes the
position. Where a deposit is incomplete the angle is wrong while the axis stays
right, and the cone is confidently misplaced rather than obviously bad ---
invisible in an image, visible in a residual. Pixel multiplicity, which governs
when a sum is complete, appears in 11 works; use of the detector as a
proportional counter, summing the full deposit to support quantification, in 7.
Ordering --- which interaction came first --- is addressed in 41, more than any
other aspect of the stage and correctly so, since an inverted pair yields the
right axis with the wrong angle.

\subsubsection{Attention follows distance from the visible result.}
Detection is examined in 76 works and image formation in 79, both directly
legible in a claim. Event building sits between them at 34, the fewest of any
stage, and it produces an object nobody sees: a better cone appears downstream
as a better image, and the improvement is credited to the reconstruction that
consumed it. \iffull   % ---- solo arXiv: sec:building, la consecuencia estructural del reparto de atencion
The consequence is structural rather than sociological. The stage
with the least attention is the one whose output every later stage inherits
without recourse, since a cone with the wrong opening angle cannot be repaired
by any inversion. Computationally it behaves as the front end does --- bounded
per-event cost once window and retention are fixed --- while being the place
where separability fails in all three of its forms at once: an ordering
constraint, a dependence on accumulated state, and arrival out of order across
block boundaries. Hard to build, cheap to run, decisive for everything after.
\fi

\subsubsection{The inverse problem moved from hardware into computation.}
\iffull   % ---- solo arXiv: sec:building, la comparacion con el colimador
A pinhole or collimated camera solves the inverse problem in hardware: the
collimator discards every photon whose direction disagrees with the geometry,
and what reaches the detector is already close to a projection. The price is
efficiency and the reward is that little remains to be computed. Removing the
collimator recovers the efficiency and moves the inverse problem into
computation, so the cost of image formation is not incidental to this design;
it is the design~\cite{feng2020benchmark}.
\fi
 Each accepted event yields a conical
surface of possible origins, constrained but not determined, and the source is
recovered where many such surfaces agree --- the information carried by the
intersection of a population rather than by any member of it. Every event is
therefore propagated as a surface against a volume, no two cones sharing a
support, so the cost of an inversion is bounded below by the event count and by
the cells each event visits, a product growing with field of view,
discretisation and opening angle.

The stage is accordingly the most abundant and most varied in the corpus: 79 of
83 works, with back projection and the MLEM family in 50 each, sparse or
algebraic formulations in 32, learned models in 27, origin-ensemble methods in
18 and analytical inversions in 16. Variety at that level marks a stage still
open, and it is the one place in the chain where an idea can be tried at the
cost of an implementation. What it does not supply is comparability: the
approaches are reported against different event counts, on different grids and
for different sources, so the choice between them cannot be made from the
literature by anyone who has not already built one.

\subsubsection{What a change of tooling cannot do.}
The difficulty here is essential in Brooks's sense~\cite{brooks1987silver}: the
propagation of a surface against a volume, once per event, following from the
geometry of the measurement rather than from any way of expressing it. Adding a
learned model can move work about and can genuinely reduce the event count a
decision needs, by selecting better or carrying a prior; what it cannot do is
make a cone stop being a surface. It arrives, moreover, with a per-event
inference time stated in none of the 83 works, so a characterised difficulty is
exchanged for an uncharacterised one --- accidental complexity added on top of
essential complexity, which is the outcome the original essay predicted for
this move. What does act on the essential term is a change of representation: a
precomputed projection operator, a sparse formulation, an algebraic sampling of
the conical support, an encoding of the event stream into a bounded angular
state~\cite{yao2020algebraic,albiol2026bounded}. These change what is
propagated and how often, so the per-event surface is paid once rather than at
every iteration. They are not faster implementations of the same computation;
they are different computations with a different lower bound, and until the
essential term is stated --- events, cells visited, and the representation that
decides how many of each --- an improvement cannot be attributed and a
real-time claim cannot be checked.

\subsection{The Image and Its Consumer}
\label{sec:handover}

The output of this chain is the input of another one. An image earns its place
by being quantifiable, either as a view a person reads or as a number a system
consumes --- a fall-off position, an uptake, an activity concentration, a
direction to avoid --- and the receiving system decides which form is
admissible. Different consumers accept different products: a scalar with a
validity stamp and a confidence, arbitrated by a safety logic; a direction and
a depth, rendered where the hands already are; a map and a threshold crossing,
tied to a position and durable enough for a record that outlives the
campaign.
What each of them accepts is simpler than what produced it, and that is a
property of the delivery rather than an accident of the consumer. A supplier
earns adoption by absorbing complexity on its own side: the front-end devices
of Section~\ref{sec:detection} are the working example, since what they present
outward --- a channel, a timestamp, an energy --- is far simpler than what they
contain, and they reached general use by narrowing what they undertake to the
level they could carry and delivering that reliably. The same discipline
applied at the far end of the chain would mean presenting a fall-off position
with a confidence, or a map with a threshold, rather than a reconstruction
together with its parameters, its calibration schedule and an integration task.
Where the technology is in service the output has that form. Where it is
proposed and not adopted, the complexity is passed outward and the receiving
party is asked to absorb it, which is a request made to whoever has least
capacity to grant it.

Three consumers, three different simplifications of the same reconstruction,
and three different values of $Q_{\min}$ --- each one a quantity the receiving
discipline sets rather than the imaging one.

The corpus is thin at exactly this junction. Of the 70 works whose full text
was searched, 42 name no downstream consumer of the image at all; the most
frequent form of handover described is an optical overlay (14), followed by a
procedure or workflow mentioned in passing (7), an operator (6) and a clinician
or surgeon (6). Four attach any number to the consumer rather than to the
computation. The clearest case in the corpus states a production rate of 75
frames per second alongside a delivery-accuracy criterion~\cite{corpus46realtime2024};
two others give a refresh cadence for a survey
product~\cite{corpus44quantitative2026,corpus55wiitm2022}. Of the 36 works set
in a therapy or range-verification context --- where the consumer is by
construction an accelerator control system --- none specifies the interface to
it: the format, the latency that system accepts, or the criterion on which it
would act. The word \emph{feedback} appears in four, in every case in the
prospective voice: the result is \emph{beneficial to}, \emph{critical for},
\emph{paves the way for}, \emph{potentially} enabling a closed loop.

\iffull   % ---- solo arXiv: sec:handover, la frontera entre dominios (la mide sec:domains)
A low count here reflects a boundary between domains rather than an absence of
the problem. The last step crosses out of instrumentation and into the industry
that operates the receiving machine, or into a clinical circle with its own
literature, its own regulatory vocabulary and its own venues. This is the same
gradient noted throughout this walk, at its far end: value accrues towards the
end of the chain, and the final handover accrues enough of it to be settled
inside product programmes and qualification dossiers rather than in the
instrumentation literature. The consequence for a review is practical --- the
acceptance criteria that would close the argument are held in a place the
citation record does not reach --- and it is the reason this stage is invoked
in 49 works and specified in far fewer.

The gap has a plain form. A measurement reaches the end of the chain and the
party receiving it cannot act on it: the quantity arrives without the
conditions, the cost or the bound that would let it be placed in an existing
procedure. At that point the result stops resolving the problem and starts
adding to it, because the work of making it usable has been transferred to
whoever received it. A demonstration that a quantity is determinable is a
genuine result; it becomes an instrument when a procedure delivers that
quantity at a stated cost, on every batch, under the conditions the receiving
system imposes. The distance between the two is what separates a phase
prototype from a deployed one, and it is measured in the terms assembled here
rather than in image quality alone.
\fi

Stating the size of that distance is the useful act. A handover has a
magnitude: the quantity delivered, the confidence attached to it, the cadence
at which it arrives, and the bound within which it is guaranteed to
arrive~\cite{albiol2026bounded}. Those four are quotable at design time, before
a prototype exists, and they are what a receiving discipline needs in order to
say whether the instrument fits its loop. The field currently supplies the
first two on occasion, the third rarely and the fourth almost never --- and the
fourth is the one that makes the other three composable with somebody else's
system.

\iffull   % ---- solo arXiv: sec:twotimes, subseccion entera
\subsection{Two Kinds of Time: Statistical and Computational}
\label{sec:twotimes}

The time an imaging problem consumes has two components of different kinds. The
first is the time for the measurement to become reliable --- enough events,
from a source of a given activity, through a geometry of a given efficiency ---
and it is a physical limit that no computation shortens, its only levers being
those of Section~\ref{sec:detection}. The second is the time to turn those
events into an image, a technological limit with the property technological
limits have: it moves. The two are addressed with very unequal effort, and in
inverse proportion to how binding they are. The technological term is where 79
of the 83 works sit and where six families of method compete; the physical term
is barely characterised, the statistics a decision requires stated in a handful
of works and the source complexity governing it swept in none. Algorithmic cost
is accordingly a diminishing part of the problem --- it remains worth reporting,
since a term that cannot be declared cannot be composed --- while the term that
decides whether these instruments reach the applications they invoke is the
physical one, and it is the term the field measures least.
\fi

% ===========================================================================
\section{Scalability and the Open Fronts}
\label{sec:fronts}
% ===========================================================================

Scaling this instrument is posed as two questions, and the two tables that
follow answer them separately. The physical table asks what limits the
measurement: its entries vary with source, crystal and geometry. The
computational table asks what a stage must satisfy for the chain to compose at
all: its entries hold whatever detector is attached, being properties of how
work is decomposed rather than of what is measured. Keeping them apart matters,
since an improvement in one is routinely offered as relief for the other.

A third question decides adoption and is answered by neither. A device is
admitted into a workflow that already exists, and what it costs there is
counted in the steps it adds to an operator whose attention is committed. \iffull   % ---- solo arXiv: sec:fronts, el recorrido del flujo de terapia
A
course of particle therapy is among the most intricate combinations of imaging
and apparatus in medicine --- planning tomography, positioning and
immobilisation, registration between modalities and again to the imaging on the
treatment machine, delineation, optimisation, robustness evaluation, plan
review, per-patient quality assurance --- assembled from several manufacturers,
each qualified separately, under a schedule that is full before any new
instrument arrives. The cost of adding a party to an arrangement of that kind
is not the work that party does but the coordination it
requires~\cite{brooks1975mythical}, and each of the disciplines above must
concede something before the next can act.
\fi
 A device therefore competes on the
improvement it delivers together with the steps it adds or removes, and the two
trade against each other. The two instruments that reached patients in this literature both
entered by adding exactly one step to a procedure already being performed.
Operational cost is as transferable as computational cost when it is written
down --- the steps required, who performs them, where they sit in a schedule,
and what they replace --- and where it is absent, each adoption decision
becomes a research project whose cost the receiving discipline pays and the
reporting discipline never sees.

The physical table is operational as well as descriptive: its columns are the
inputs to a calculation every group performs and few write down. Detector
uncertainties fix the smallest cell that carries information; the intended
extent fixes how many such cells the region contains; the events per resolvable
element then fix how long the measurement must run. In practice the reasoning
is carried out informally, by inspection --- our own grids have been sized that
way --- which is legitimate, while writing it down is what makes it
transferable: a grid arrived at by inspection cannot be carried to a detector
with different uncertainties, and neither can the time and resolution that
depend on it.

\begin{table*}[t]
\centering
\caption{Physical fronts of the acquisition. For each regime, the term that
binds first, the stage at which it binds, the answer the literature usually
proposes, what that answer costs elsewhere in the chain, and the quantity that
would be needed to judge it. These are the quantities that move $Q_{\max}$ and
the window; the computational properties that decide whether a chain can be
composed at all are the subject of Table~\ref{tab:comp}.}
\label{tab:fronts-phys}
\footnotesize
\begin{tabular}{@{}p{0.105\linewidth}p{0.145\linewidth}p{0.10\linewidth}p{0.155\linewidth}p{0.20\linewidth}p{0.205\linewidth}@{}}
\toprule
\textbf{Front} & \textbf{What binds first} & \textbf{Where} &
\textbf{Usual answer} & \textbf{What it costs} & \textbf{What is not reported} \\
\midrule
Low rate &
Statistics: $N_k/R_{\mathrm{acc}}$ exhausts the window before the image forms &
Geometry and efficiency &
Larger or closer detectors, more modules, longer acquisition &
Proximity degrades the projection; more modules multiply channels and intrinsic
activity; longer acquisition breaks the window &
The trade of solid angle against near-field projection error; the intrinsic
count rate that arrives with more crystal \\
\addlinespace
High rate &
Acquisition: per-channel ceiling, dead time, buffer overflow &
Front-end and transport &
Faster ASIC, early filtering in firmware, narrower coincidence windows &
Early reduction may be irreversible; loss becomes block-structured &
Loss attribution, live time, and the interval each loss covers \\
\addlinespace
Low resolution &
Nothing binds; this is the regime that works &
--- &
Coarse cells, direct back projection, few iterations &
A quality ceiling that may fall below what the task requires &
That it is a deliberate operating mode rather than a degraded result \\
\addlinespace
High resolution &
Cost per event times cells visited, while events per cell falls &
Image formation &
Accelerator, precomputation, sparse operator, learned model &
Setup and inference time enter the loop; memory grows; dispersion per cell
rises &
The cost of the precomputation, the inference time per event, and the
events-per-cell regime in which the claimed resolution holds \\
\addlinespace
High source entropy &
Both terms at once: $N_k$ grows with the degrees of freedom, and event-wise
inversion degrades &
Acquisition and image formation &
More events, more iterations, regularisation, learned priors &
Time; and a learned prior changes what the image asserts about the source &
Any sweep of source complexity; the normalisation of event counts by
resolvable source element \\
\addlinespace
Low source entropy &
Nothing binds; the point source is the favourable case &
--- &
The default test object of the field &
Results obtained here do not transfer upward &
That the reported figure is conditioned on the easiest source \\
\bottomrule
\end{tabular}
\end{table*}

\begin{table*}[t]
\centering
\caption{Computational fronts. Where the physical table asks what limits the
measurement, this one asks what a stage must satisfy for the chain to be
composed at all. These properties are independent of the source and of the
crystal: they hold or fail for the same reasons at every rate and every
resolution. The final column gives the count from the corpus of
Section~\ref{sec:survey}.}
\label{tab:comp}
\footnotesize
\begin{tabular}{@{}p{0.11\linewidth}p{0.175\linewidth}p{0.17\linewidth}p{0.165\linewidth}p{0.155\linewidth}p{0.115\linewidth}@{}}
\toprule
\textbf{Property} & \textbf{What it requires} & \textbf{Where it fails} &
\textbf{Usual answer} & \textbf{What it costs} & \textbf{Reported} \\
\midrule
Bounded state &
Steady-state memory independent of acquisition length &
List-mode accumulation, whose state is the run itself &
Snapshots, histograms, ordered subsets &
A change of representation, and a loss of per-event provenance &
Memory footprint in 9 of 83 \\
\addlinespace
Bounded latency &
Every stage declarable from configuration and verifiable by measurement &
Any stage whose cost depends on accumulated history, or that inherits a
filesystem tail &
Change of representation; keep storage off the flow-control path &
A setup cost paid once, which then has to be declared &
Cost of precomputation in 0 of 83 \\
\addlinespace
Separability &
Work items independent: no ordering constraint, no shared reduction, no
state-dependent decision &
Grouping fails on ordering, coincidence on state, image accumulation on the
shared reduction &
Partition by detector; per-worker accumulation with a bounded merge &
The merge becomes a measured quantity rather than an assumed one &
Discussed in 28 of 83 \\
\addlinespace
Amortisation &
A one-time cost distinguished from a per-event cost &
System matrices, inversion kernels and learned models: setup paid once, use
paid always &
Precompute or train once, reuse thereafter &
The per-event term enters the latency budget and may exceed the arithmetic it
replaced &
Inference time in 0 of 83; training cost in 5 of 37 \\
\addlinespace
Composability &
The output of a stage reaches the next without passing through a file &
Store-and-reprocess, where storage is the channel between stages &
Streaming, with recording as a tap rather than an interface &
Provenance must be carried explicitly, which filenames used to supply &
Storage absent from every reported budget \\
\addlinespace
Observability &
Each stage reports what it lost, when, and over what interval &
Loss detected downstream as a gap, with the originating stage unrecoverable &
Counters carrying sequence range and timestamp interval &
A small permanent bookkeeping overhead &
Loss interval and live time essentially unreported \\
\bottomrule
\end{tabular}
\end{table*}

\subsection{Source Complexity, and What Makes a Count Comparable}
\label{sec:entropy}

The physical table sizes a measurement; one term in that sizing carries a
dependence that is almost never declared, and it is the term that decides
whether a published figure can be carried anywhere at all.

The events a decision requires are not a property of the reconstruction method
alone but of the method together with the source, and the governing property of
the source is its spatial entropy. A point-like emitter concentrates its
evidence, so few lines determine it; an extended or structured distribution
spreads the same evidence over many degrees of freedom and requires
correspondingly more. Written explicitly,
\begin{equation}
 N \;=\; N\bigl(H_{\mathrm{source}},\, \text{confidence},\, \text{method}\bigr),
\label{eq:entropy-dependence}
\end{equation}
which makes visible what a bare event count conceals. The dependence does not
stop at acquisition: a distribution of higher entropy activates more of the
sparse support during inversion, so $H_{\mathrm{source}}$ enters the
reconstruction cost as well. Source complexity therefore appears in both terms
of the decision time --- in the measurement needed to reach a given quality,
and in the computation needed to use it --- and a quoted event count without a
stated source complexity is uninterpretable rather than merely incomplete.

The normalisation follows directly. Taking $\delta$ as the smallest cell the
measurement supports, derived from the propagated detector uncertainty as
above, and $D$ as the extent of the activity, the source occupies
\begin{equation}
 M_{\mathrm{source}} \approx \left(\frac{D}{\delta}\right)^{d},
 \qquad
 N_{\mathrm{eff}} = \frac{N}{M_{\mathrm{source}}},
\label{eq:source-dof}
\end{equation}
so that a point source occupies one element and an extended distribution
occupies many. Two figures quoted for different sources become comparable only
after each is divided by its own $M_{\mathrm{source}}$. This is the mirror, on
the source side, of the events-per-cell normalisation the image side already
imposes: one quantity fixes what the measurement can resolve, the other what it
must resolve, and a claim about capability needs both.

Two consequences follow. Event counts from different works are not comparable
unless the sources are, so the ratio between two such figures measures the
sources at least as much as the algorithms. And source complexity is not only a
scaling of cost: a method benign on a concentrated source can fail on a
distributed one for reasons of conditioning rather than resources, so a result
demonstrated at low $H_{\mathrm{source}}$ does not extend to high
$H_{\mathrm{source}}$ by adding events or machines.

This is the technical half of transferability, and it is measured as thinly as
the rest. Source complexity is never swept in the corpus
(Fig.~\ref{fig:corpus}): extended or distributed activity appears in 10 works of
83 and in several of those as a statement of intent, phantoms and point sources
are the default test objects, and of the occurrences of the word entropy none
names a property of the source --- they name a loss function in a learned
method. A field that reports counts without $M_{\mathrm{source}}$ has published
numbers that compare only to themselves.

\subsection{What a Cost Statement Has to Name}
\label{sec:costdecomp}

The entries of the computational table are shorthand, and shorthand is a poor
instrument for a cost. A statement such as ``the reconstruction is
$O(N\,M)$'' is complete only when it names the resource that runs out first,
and the resources run out separately.

\iffull   % ---- solo arXiv: sec:costdecomp, los cuatro recursos uno a uno
\emph{Arithmetic} is the term usually quoted, and it decomposes further than
one symbol allows: a cost per event, a cost per cell touched by that event, a
cost per iteration, and a number of iterations. The last of these is governed
by the statistics and by the source, so it varies with the same quantity that
Section~\ref{sec:regimes} uses to normalise the measurement; an iterative
method quoted at a fixed iteration count has been quoted at one operating
point.

\emph{Memory} is what precomputation buys with, and it is the resource most
often spent silently. A system matrix, a sensitivity map, an efficiency table
per pixel or per detector pair: each converts arithmetic into residency, and
residency is bounded by a device rather than by patience. The relevant figure
is the working set at the operating point rather than the total size of the
table, because a table larger than memory is a different kind of object.

\emph{Access} is the term that appears once a table stops fitting, and it is a
database question rather than an arithmetic one. An efficiency matrix consulted
once per event, with an index that follows the events rather than the storage
order, is a random-access workload whose cost is set by locality and by the
layout chosen for the table. A method can be arithmetically cheap and
access-bound, and the two are not distinguishable from a complexity class.

\emph{Synchronisation} is what a parallel decomposition costs, and it is
carried by the barriers rather than by the work between them. A map-reduce
formulation is described by where its joins fall, what must be complete before
each one, and what the merge costs when the partitions are uneven. The three
failure modes of separability --- an ordering constraint, a shared reduction, a
state-dependent decision --- are three ways of placing a barrier, and naming
the barrier is more useful than naming the mode: a stage that joins once per
acquisition block and a stage that joins once per event have the same
separability description and different behaviour under load.
\else
% BRIDGE (texto nuevo, solo version revista)
Four resources have to be named separately --- arithmetic, memory, access and
synchronisation --- because they run out separately, and a complexity class
names none of them.
\fi

These four exchange against one another rather than adding. Precomputation
converts arithmetic into memory and access; partitioning converts wall time
into synchronisation; a change of representation moves the whole problem from
one resource to another, which is why it changes the lower bound where a faster
implementation does not. Which resource binds is a property of the operating
point and not of the method, and the same algorithm is compute-bound at one
rate and access-bound at another. A cost quoted without its operating point ---
the event rate, the source extent, the grid, the residency available --- states
which resource was abundant on the machine that ran it, and a reader with a
different machine and a different source learns little that transfers.

% ===========================================================================
\section{What the Literature Reports}
\label{sec:survey}
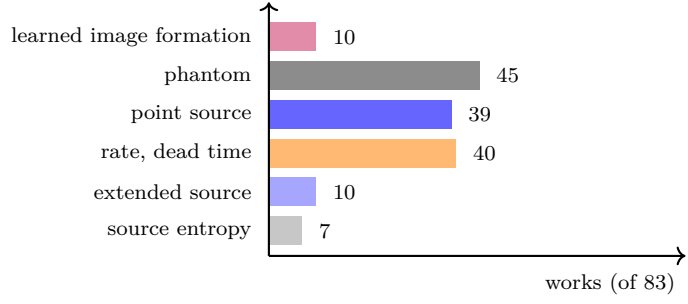
\begin{figure}[t]
\centering
\begin{tikzpicture}[scale=0.95]
\def\bar#1#2#3#4{%
  \fill[#4] (0,#1) rectangle (#2/83*5.4,{#1+0.40});
  \node[anchor=east,font=\scriptsize] at (-0.1,{#1+0.20}) {#3};
  \node[anchor=west,font=\scriptsize] at ({#2/83*5.4+0.1},{#1+0.20}) {#2};}
\bar{0.10}{7}{source entropy}{black!22}
\bar{0.64}{10}{extended source}{blue!35}
\bar{1.18}{40}{rate, dead time}{orange!55}
\bar{1.72}{39}{point source}{blue!60}
\bar{2.26}{45}{phantom}{black!45}
\bar{2.80}{10}{learned image formation}{purple!45}
\draw[->,thick] (0,-0.05) -- (0,3.48);
\draw[->,thick] (0,-0.05) -- (5.8,-0.05);
\node[font=\scriptsize,anchor=north east] at (5.8,-0.14) {works (of 83)};
\end{tikzpicture}
\caption{The image and source axis. Phantoms and point sources are the default
test objects; extended or distributed activity appears in 10 works, and in
several of those as a statement of future intent. Source complexity is never
treated as a variable to be swept, and no occurrence of the word entropy names
a property of the source. Ten works form or enhance the image with a learned
model, which places a further undeclared term between the measurement and the
picture.}
\label{fig:corpus}
\end{figure}

% ===========================================================================
Three counts follow. The first two describe what the literature of this
instrument reports about the source it images
(Fig.~\ref{fig:corpus}) and about its own computation
(Fig.~\ref{fig:corpus-algo}); the third asks where that instrument is wanted
and where it is used.
The corpus, and the screening that produced these counts, are described first,
so that every count arrives with the procedure that generated it.

\subsection{How the Corpus Was Assembled}
\label{sec:method}

Records were exported from a single bibliographic database using title,
abstract and keyword queries over Compton and gamma-ray imaging, its
acceleration and parallelisation, its acquisition and readout, and the
applications that motivate real-time operation. That export was screened by
hand: a record was retained where its subject was such a system, its
reconstruction, or its readout, and discarded otherwise along with duplicates
and correction notices. To the result we added works cited by it that supply
terms the argument needs, so that the sample is not confined to what a keyword
search surfaces. Eighty-three full texts were obtained and are the basis of
every count. Denominators differ between counts because not every full text
supports every analysis: 83 works were scored on subject markers, 70 yielded
text clean enough for the operating-point and reporting markers, 58 had
reference lists that could be parsed, and 42 of those could also be assigned a
stage. Each count states its own denominator, and none is a subset chosen after
the result was known; all appear in the bibliography, and the marker definitions
together with the per-work screening are supplied as supplementary material, so
that any figure given here can be recomputed or contested at the level of the
individual record.

Two precautions were taken against selection bias. The works discussed
individually were chosen to span groups and countries rather than to make a
point about any of them, and our own companion work is included in the corpus
and scored by the same markers as everything else, with the result reported
whether or not it is favourable. A word on the zero counts, since several carry
weight: the survey was run twice, on a smaller corpus and then on the present
one after duplicates were removed and further records added. Between passes the
markers that matter moved by less than two percentage points, while the counts
that were zero remained zero as the corpus grew.

\subsection{Counting in Context: What Verification Changed}
\label{sec:verify}

The counts reported here come from full texts read in context rather than from
keyword matches, and the distinction is worth setting out because in several cases it
changed the result. A marker was accepted only where its occurrence carried the
meaning the count claims, and the occurrences were inspected to establish that.

Three examples show what the inspection removes. Searching for a declared
purpose returns the phrase \emph{intended use} in seven works; all seven
instances are the Creative Commons licence footer, and the count in the sense
meant here is zero. Searching for a regulatory pathway returns two matches, and
both are voltage regulators. Classifying works by application using the highest
marker count placed fifty of seventy in nuclear medicine, an artefact of a
generic vocabulary appearing in passing throughout the corpus; repeating the
classification on title and abstract, where an application is actually
declared, gives twenty. Smaller corrections run the same way: most occurrences
of \emph{adapt} in the range-verification works are \emph{adapted},
\emph{adapters} or the adaptation of a method rather than of a treatment, and
of five works matching a clinical-trial marker, the two describing patients
under measurement use time-of-flight PET and a freehand SPECT probe, both
established modalities rather than the instrument under review.

The corpus itself is listed rather than described. Every work counted appears
in the bibliography, including those not discussed individually in the text,
so that any figure given here can be recomputed by a reader who obtains the
same documents. A count over a corpus its author does not enumerate cannot be
checked, which is the property this review asks of the works it surveys and
therefore owes them.

The direction of these corrections is uniform and is the reason they are
reported. Every one reduced a count: a marker that appeared present proved
absent on inspection, and no inspection converted an absence into a presence.
Each figure given here is therefore an upper bound on how much of the field
reports the property in question, and the zeros are conservative --- a work
would have had to state the quantity in words no reading of its full text
recovered. Counts assembled from titles, abstracts or keyword indices alone are
higher than these, and the difference between the two is itself a measure of
how much of the reporting is nominal.

One limit on how these counts should be read follows from what they are. They
record what a reporting convention has produced up to a stated date, not
whether an author met a requirement --- no such requirement was in force when
the works were written, and none is imposed on them retrospectively here. The
quantities this review asks for are offered for what is reported next. A work
that does not state its operating point was not failing a standard; it was
following the convention of its field, which is the object under study.

\begin{figure}[t]
\centering
\begin{tikzpicture}[scale=0.95]
\def\bar#1#2#3#4{%
  \fill[#4] (0,#1) rectangle (#2/83*5.4,{#1+0.36});
  \node[anchor=east,font=\scriptsize] at (-0.1,{#1+0.18}) {#3};
  \node[anchor=west,font=\scriptsize] at ({#2/83*5.4+0.1},{#1+0.18}) {#2};}
\bar{0.10}{0}{cost of precomputation}{black!22}
\bar{0.58}{0}{complexity notation}{black!22}
\bar{1.06}{8}{stated scaling law}{black!30}
\bar{1.54}{13}{computational complexity}{green!25!black!40}
\bar{2.02}{9}{memory footprint}{green!35!black!50}
\bar{2.50}{28}{separability discussed}{green!45!black!60}
\bar{2.98}{16}{precomputation used}{green!50!black!70}
\bar{3.46}{39}{accelerator used}{green!55!black!80}
\bar{3.94}{31}{learned model used}{purple!55}
\bar{4.42}{0}{ML inference time}{black!22}
\draw[->,thick] (0,-0.05) -- (0,5.05);
\draw[->,thick] (0,-0.05) -- (5.8,-0.05);
\node[font=\scriptsize,anchor=north east] at (5.8,-0.14) {works (of 83)};
\end{tikzpicture}
\caption{The algorithmic axis, and an inversion. The mechanisms that produce
the reported speed are widely used --- an accelerator in thirty-nine works,
precomputation in sixteen --- while the properties that would let a reader
carry the result to another system are not. Separability is discussed in
twenty-eight, computational complexity in thirteen, a scaling statement made in
eight, a memory footprint given in nine, and the cost of the precomputation
reported in none.}
\label{fig:corpus-algo}
\end{figure}
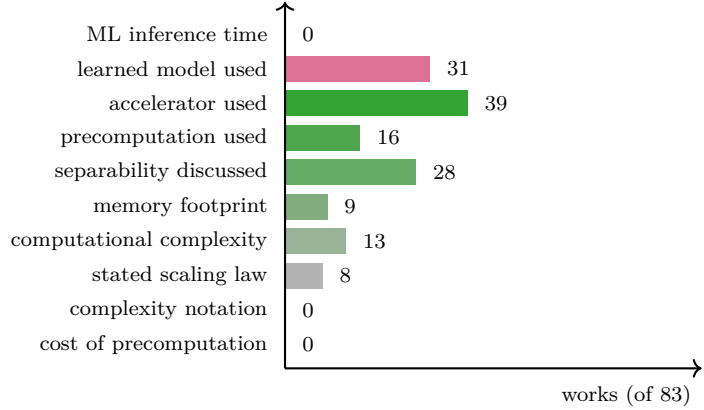

\begin{figure}[t]
\centering
\begin{tikzpicture}[scale=1.12]
\def\pb#1#2#3#4{%
  \fill[blue!55!black] (0,{#1+0.20}) rectangle ({#2/100*4.3},{#1+0.38});
  \fill[orange!75!black] (0,#1) rectangle ({#3/100*4.3},{#1+0.18});
  \node[anchor=east,font=\scriptsize] at (-0.08,{#1+0.19}) {#4};
  \node[anchor=west,font=\scriptsize] at ({#2/100*4.3+0.06},{#1+0.29}) {#2};
  \node[anchor=west,font=\scriptsize] at ({#3/100*4.3+0.06},{#1+0.09}) {#3};}
\pb{5.15}{87}{38}{spatial resolution, FWHM}
\pb{4.65}{37}{9}{reconstruction or compute time}
\pb{4.15}{58}{66}{phantom or test object}
\pb{3.65}{5}{61}{clinical endpoint}
\pb{3.15}{10}{61}{patients treated or imaged}
\draw[gray!55,dashed] (-2.35,2.98) -- (5.3,2.98);
\node[font=\scriptsize,anchor=west,gray!45!black] at (-2.33,2.83)
      {\itshape quantities that would let either result be carried to the other};
\pb{2.30}{1}{0}{operating point declared}
\pb{1.80}{0}{0}{cost per event}
\pb{1.30}{0}{0}{named user of the output}
\pb{0.80}{0}{0}{result re-derivable later}
\pb{0.30}{0}{0}{numerical determinism}
\pb{-0.20}{0}{9}{transferability discussed}
\draw[->,thick] (0,-0.28) -- (0,5.72);
\draw[->,thick] (0,-0.28) -- (5.0,-0.28);
\node[font=\scriptsize,anchor=north east] at (5.0,-0.37) {\% of corpus};
\fill[blue!55!black] (2.5,5.62) rectangle (2.78,5.76);
\node[font=\scriptsize,anchor=west] at (2.82,5.69) {instrumentation, $n{=}70$};
\fill[orange!75!black] (2.5,5.40) rectangle (2.78,5.54);
\node[font=\scriptsize,anchor=west] at (2.82,5.47) {clinical, $n{=}20$};
\end{tikzpicture}
\caption{The clinical and the instrumentation literature, measured on the same
markers. Above the line each corpus is rich in its own quantities and sparse in
the other's, which is expected and unobjectionable: a resolution figure appears
in 87\% of instrumentation works and a clinical endpoint in 61\% of clinical
ones. The phantom is the single object both discuss at comparable frequency,
58\% against 66\%, which is why metrology is the natural place for the two to
meet. Below the line are the quantities that would allow a result from either
to be used by the other, and both corpora are empty of them. Consistently, of
the 58 instrumentation works whose reference lists could be parsed, citations
to venues in which clinical outcomes are reported number 47 against 971 to
instrumentation venues, and 34 cite no such venue at all. The two markers are
defined sets rather than a partition: medical-physics venues belong to
neither.}
\label{fig:junction}
\end{figure}
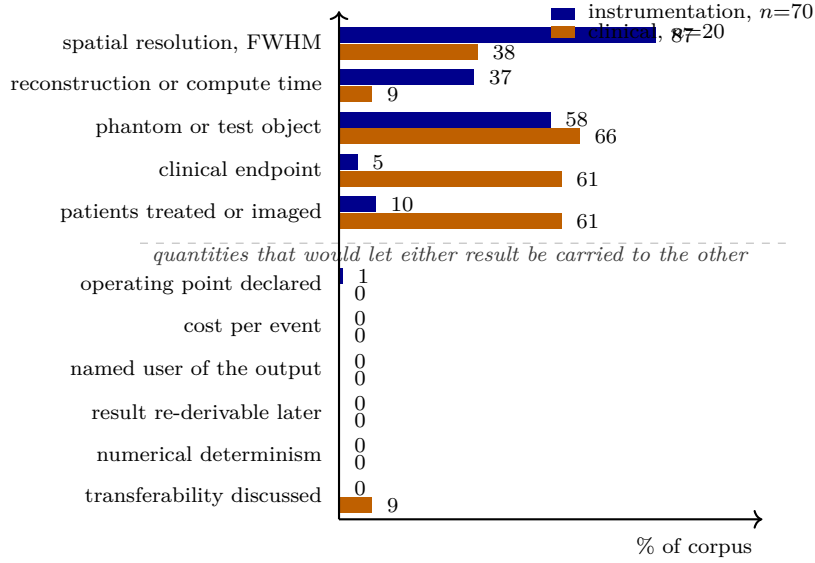

\subsection{Application Domains: Where the Capability Is Wanted and Where It Is Used}
\label{sec:domains}

The counts above describe what the literature of this instrument reports. A
second question needs a different measurement: whether a capability
demonstrated in one application can be pointed at another, and what governs
that. Five application domains were measured on one template ---
the size of the community that would consume the image, whether that community
states an unmet capability in its own literature, what instrument serves the
task today, the size of the Compton literature aimed at the domain, and the
resources the community has committed towards deployment. Table~\ref{tab:domains} gives the
result.

These figures come from record counts in a single bibliographic database,
without the manual screening applied to the full texts of
Section~\ref{sec:verify}, and they are reported as a distinct and weaker class
of evidence for that reason. They support statements about relative scale and
about ratios between domains measured identically; they do not support
statements about individual works.

\begin{table*}[t]
\centering
\caption{Five application domains on one template. \textbf{Destination} is the
literature of the community that would consume the image; \textbf{need}, that
part of it concerned with the capability this instrument would supply, which
is narrower than the field's largest problem; \textbf{incumbent}, the
instrument serving the task today; \textbf{ours}, the Compton literature aimed
at the domain. \textbf{Committed} counts works in which the
community has committed resources towards deployment, and its definition
differs by domain: for astrophysics, flight hardware, a balloon campaign or a
selected mission; for particle therapy, a clinical endpoint; for
decommissioning and survey, field, on-site or vehicle-borne measurement. The
column does not measure service, and the difference matters in both directions.
The decommissioning works of the full-text corpus describe themselves as
prototypes, proofs of concept and feasibility demonstrations, and report a
decision or action taken as a result in essentially none. On the astrophysics
side no Compton telescope has operated in orbit since COMPTEL ceased returning
data, the selected missions counted here have not yet flown, and one that did
was lost early; that side could not be checked in full text because those works
are outside the corpus, so the asymmetry of scrutiny runs against the reading
we would prefer. Environmental survey was measured and discarded, its query
admitting an ordinary English word. Rows are ordered by the last column.}
\label{tab:domains}
\small
\resizebox{\textwidth}{!}{%
\begin{tabular}{@{}lrrrrrl@{}}
\toprule
\textbf{Domain} & \textbf{Destination} & \textbf{Need} & \textbf{Need/Dest.} &
\textbf{Incumbent} & \textbf{Ours} & \textbf{Committed} \\
\midrule
Decommissioning, survey & 1981 & 58 & 2.9\% & 44 & 99 & 71 in the field (72\% of ours) \\
Astrophysics & 1600 & 136 & 8.5\% & 4500 & 219 & 130 on a flight track (59\% of ours) \\
Particle therapy & 4000 & 600 & 15\% & --- & 128 & 8 with a clinical endpoint (6\%) \\
Security, safeguards & 2500 & 73 & 2.9\% & 227 & 111 & --- \\
Nuclear medicine & 41000 & 148 & 0.4\% & --- & 128 & --- \\
\bottomrule
\end{tabular}
}
\end{table*}

Four explanations are available, three before the data are consulted and one
after, and the data remove all four. The size of the destination does not
discriminate: the four measured destinations span 1600 to 41\,000 and the
extremes sit at opposite ends of the last column. Nor does the size of the
incumbent: the largest measured belongs to the domain with the most commitment.
Nor does the absolute size of a stated need: astrophysics and nuclear medicine
state needs of 136 and 148 records with opposite outcomes. Nor --- against our
own expectation, which we record because it was wrong --- does that need
relative to its destination: particle therapy states the largest, 15\%, and
shows the least commitment, while decommissioning states 2.9\% and shows the
highest. The volume of a stated need measures how much a discipline discusses a
limitation rather than how far it is from acting on one.

Two properties survive, and neither belongs to the instrument. The first is
whether the incumbent can serve the task at all. An instrument filling a gap
its predecessor cannot reach by construction --- a band no telescope covers, a
volume no collimated survey enters without discarding the photons that carry it
--- competes against nothing, and both domains with the strongest deployment
commitment are of that kind. An instrument offered against a task an incumbent
already performs must displace it, which requires different evidence on a
different timescale; the two domains with the least commitment face, respectively,
margins with computed-tomography-based range prediction and an installed,
reimbursed modality. Filling a gap and displacing an incumbent are routinely
presented as one proposition and are not comparable in cost, evidence or
duration.

The second is how many domain boundaries the output must cross before it is
acted on. In astrophysics the community that consumes the image is the one that
builds the instrument, so the chain closes without a handover. In site
characterisation the output crosses one boundary, to an operator who accepts a
map with a threshold. In particle therapy it crosses three --- a manufacturer
who owns the delivery system, a regulator who qualifies it, and a clinical
service whose schedule is committed --- and each must concede something before
the previous one can act. That count orders the last column ---
one boundary, none, three --- more cleanly than along any property of the
detector or of the receiving literature. Three domains ordering consistently is
a regularity rather than a law. What the count expresses is Q1 asked at the
scale of an application: not only which decision the image supports, but how
many parties must each take their own before anyone acts.

The consequence is visible in what the field publishes. Of the works whose
application is declared in title or abstract, 21 address particle therapy and
20 nuclear medicine, against 3 for decommissioning and field survey and 2 for
security and astrophysics, while evidence of deployment commitment is
concentrated in the last two groups: compact-camera prototypes on crawler
robots inside reactor buildings and on drones for prompt survey, cameras
deployed around a release to reconstruct a plume~\cite{corpus44quantitative2026},
a rotating omnidirectional camera for contamination
imaging~\cite{corpus19development2025}, a portable gamma-neutron
instrument~\cite{corpus49simultaneous2024}, a survey
camera~\cite{corpus10a2020}, diagnostics permanently installed on a fusion
machine~\cite{corpus20efdc1405092012}, and balloon and satellite instruments.
The distribution of publication is close to the inverse of the distribution of
deployment commitment. These records do not establish routine service: the
decommissioning cases remain prototypes, and the astrophysical cases record a
scientific mission pathway, whose historical feasibility was demonstrated by
COMPTEL.

% ===========================================================================
\section{Discussion}
\label{sec:discussion}
% ===========================================================================

\subsection{Conditional Claims: What Follows ``Could''}
\label{sec:could}

Claims about future benefit are made in the conditional throughout this
literature, which is appropriate: the work is prospective and the mood matches.
What separates a useful claim from a decorative one is the object the
conditional takes, and four statements from the corpus, ordered by that
criterion, make the distinction better than an argument would.

At one end it takes an adjective: imaging that ``may bring several improvements
in nuclear medicine and provide clinically exploitable images'', a system with
``the potential to be utilised in clinical imaging'', a method with ``the
potential to contribute to clinically feasible'' imaging. Each is defensible
and none can be checked, because nothing follows the modal that a later reader
could measure against.

Further along it takes an architecture: prompt-gamma simulations ``could be
precalculated for each 4D CT phase after plan approval''~\cite{berthold2026fourd},
naming the mechanism, the moment it would run and the cost it would move.
Further still, an output specification: ``a field-wise classification would be
sufficient; that is, after the delivery of each treatment field, the PGI system
would either state a relevant or a non-relevant treatment
deviation''~\cite{berthold2023detect}, with the operating point bounded in the
same work --- false positives ``should, for practical reasons, not exceed more
than roughly 20\% of all PGI triggers''. Together those are the corpus's most
complete answer to Q1 and Q2: a named decision and a stated interval in which
it must be available.

At the far end it takes a limit, and the best example is negative. Range
verification at therapeutic dose with two-millimetre precision ``on a
single-spot basis seems unfeasible, and on a plane basis would require an
increase of the setup efficiency of 15--40 times, depending on the beam ion
species''~\cite{corpus50springer20202021}. That is the most useful sentence
about future capability in the corpus and it promises nothing: it converts an
aspiration into a factor, names the quantity the factor applies to, and states
the condition under which the aspiration fails. It is also the only statement
answering Q4 in the form Q4 requires --- not what the source is, but how the
required statistics scale from the demonstrated condition to the intended one.
A related ceiling is stated by the group that would profit from overstating it:
the margin for range uncertainty ``should not be reduced further than the
intrinsic uncertainty of the PGI system''~\cite{bertschi2023margin}. The
distance between the first group of statements and these is the distance this
review asks the field to close.

\subsection{Cross-Citation Between Stages}
\label{sec:nocite}

Two things hold at once here. The research is genuinely difficult, every stage
of Section~\ref{sec:chain} carrying an open problem worked by people good at
it; and the set of applications realised is small. The second is the surprising
one, and it invites a mechanism rather than a lament.

The mechanism is visible in the reference lists. Of 42 works whose stage and
bibliography could both be read, the acquisition group cites its own stage in
11 of 13 cases and the reconstruction literature in 2; the reconstruction group
cites its own in 13 of 16 and the acquisition literature in 2. Works positioned
at the application cite most widely --- 5, 7 and 9 across the three stages ---
so within this corpus the blindness is directional: proximity to the
application obliges a work to look upstream, while distance from it removes the
obligation to look down. Each
stage delivers into the next on the understanding that the next will absorb
what arrives.

A second count answers the obvious objection, that a detector paper has little
occasion to cite an oncology outcome study. Searching the same lists for the
technical literature of verification --- quality assurance, commissioning,
dosimetry audits and intercomparisons, traceability, codes of practice and the
task-group reports through which a new measurement is admitted into practice
--- returns 27 citations against 1616 to instruments and algorithms, with 15 of
58 works citing such a title even once. That is not a distant literature
written for another purpose: it is the technical output of the discipline that
would have to accept the measurement, and the mechanism by which it admits one.

The cost is a confrontation that never takes place: resolution, time and image
quality are each reported inside the stage that produces them, against that
stage's previous version and in that stage's units.

The clearest instance is the field's own summary of its most active stage.
Kim and Lee's review of Compton camera image
reconstruction~\cite{kim2024review} is the reference a newcomer to that stage
should read first, and its scope is also a measurement. Speed is discussed
throughout, and where it is quantified it is reported as a relative speed-up
rather than as an absolute time: $9.5$ for ordered subsets against maximum
likelihood at a fixed iteration count, four in processor time for a stochastic
origin ensemble. Elsewhere the improvement is qualitative --- a learned method
that ``significantly reduced the computation time''. No absolute reconstruction time usable as a term in
a latency budget appears, and the words latency, inference time, memory,
storage and precomputation do not occur in the text; acquisition architecture, readout constraints and
workflow are not treated as objects of analysis. A ratio is the correct unit inside a
stage, because it answers which of these algorithms to use. It composes with
nothing: two speed-ups against different baselines cannot be added, placed in a
budget, or compared with the five minutes a receiving discipline has after a
field is delivered. Read against Section~\ref{sec:rank}, a ratio is silent on
Q2, naming no duration.

The same review carries the deeper consequence in its own structure. Its
section on advantages opens by naming energy resolution and sensitivity as what
the technique offers, supported by a comparison with a collimated camera; its
survey of designs closes by naming energy resolution and sensitivity as the
challenges that persist for practical application. Both statements are correct
and concern the same two quantities. What changes is the reference frame: the
first measures against a peer instrument and answers whether this technique
beats that one, the second against a task and answers whether it suffices. The
two answers are independent, and nothing in the reporting convention requires
an author to mark which frame is in use --- Q3 going unasked. This is what
allows a field to remain superior and insufficient at once for a long time.
Superiority over a predecessor is real and accumulates; sufficiency for an
application is a different statement that accumulates separately or not at all,
so a programme measured only in the first frame can advance for decades while
the second stays where it was, and the advance is genuine throughout. Binding a
quantity to a task is what makes the frames commensurable, and it is the
operation this review asks for under several names: an intended use, an
operating point, a declared cost, a stated bound.

The reading is not ours alone. The authoritative review of in-vivo range
verification frames the need as the unexploited potential of a delivery
technique rather than as a capability its users ask for, locates a major source
of the uncertainty upstream in the calibration of the planning tomograph, and
concludes that no routine solution has been established and that ``none of
them will offer alone the ultimate solution to the problem of range
uncertainty''~\cite{parodi2018invivo}. Each of those statements is made once,
in a summary, and none is composed with the others. Nor are they taken up: the
review is cited by 5 of the 83 works, in every case as a pointer in a reference
list, and no work in the corpus engages with its conclusions in its body text.
The field's own statement that no routine solution exists and that no single
technique will suffice is available, and unused.

Taken together those statements are the position this review measures: a
capability offered against a problem whose root the same text places elsewhere,
by an assembly of partial techniques none of which suffices, into a procedure
that must absorb all of them. The absorption is the difficulty. Where in-vivo
verification has come closest to routine use, the measured signal is compared
against a Monte Carlo reference computed on the patient's own anatomy, which
took hours in the reported study against a target of about five minutes after a
field is delivered~\cite{berthold2026fourd}. The instrument therefore delivers
not a quantity but a comparison, and the comparison carries a second model with
its own inputs, its own uncertainty and its own interpretation --- introduced
at the point where a patient is present and where the schedule is least able to
absorb it.

Reviews inherit the same boundary, which is why so few reach the handover of
Section~\ref{sec:handover}: a survey organised by stage stops where its stage
stops, and the last stage --- what is delivered, to whom, in what form, within
what bound --- belongs to no stage and therefore to no survey of one.

The constructive form of this is a demand for metrology rather than for
goodwill. Stating the limits of a technology for an application it has not yet
served requires quantities defined so that they survive crossing a stage
boundary: a cost bound to its operating point, a resolution bound to the source
that produced it, a latency bound to the state it was measured in, an output
bound to the consumer that accepts it. Quantities of that kind compose, and
what composes can be confronted --- which is the difference between a research
programme and a catalogue of demonstrations.

\subsection{Metrology as the Available Move}
\label{sec:metrology}

Every remedy identified above depends on somebody else: a correction applied
during delivery requires the party owning the machine to requalify it, adoption
in an operating theatre requires a committed schedule to make room, a change to
an accepted tolerance requires clinical evidence on another timescale. One
remedy depends on nobody, and the corpus shows it lying open.

The reference object is the single point at which the two literatures already
meet: a phantom or test object appears in 45 of the 83 instrumentation works and
in 66\% of the clinical ones (Fig.~\ref{fig:junction}). What is absent is the
shared definition. A named or standardised test object appears in 14 of the 83
works, an intercomparison or round-robin protocol in 2, traceability or a
reference standard in 3, and a publicly available dataset in 2. The community measures against
reference objects and builds a different one each time, so no two reported
figures are commensurable and no procedure exists by which they could become
so.

Occupying that gap carries no regulatory exposure and no clinical risk, since
its subject is an object rather than a patient, and it requires no vendor to
modify a machine and no service to alter a schedule. Its returns are also of an
unusual kind. An instrument is superseded by a better instrument; a definition
used to compare instruments is superseded only by a better definition, and
definitions of that sort turn over across decades rather than product cycles.
The precedent is close at hand: the geometric calibration literature of
radiation therapy amounts to some five hundred records sitting beneath four
thousand on clinical outcome, and it entered practice by supplying a
measurement with a reference and a procedure rather than by asserting that
existing practice was inadequate.

The order matters and is not a play on words. A methodology --- how results are
compared, what a claim must contain, which conditions accompany a figure ---
can be written only once the quantities it manipulates are defined and their
references are traceable. Metrology is therefore the prior step and the one
available now, and the terms this review has been assembling in fragments ---
an operating point, a declared cost, a bound, a named consumer --- become
writable the moment the quantities beneath them are fixed to something other
than each group's own apparatus. The transferable product of such work is a
characterised term: a quantity, its uncertainty, the conditions under which it
holds, and the interval over which it was verified. A term of that form can be
placed by each party beside its own tolerance without conceding anything to the
others, which a recommended value cannot.

\section{Conclusions}
% ===========================================================================

\begin{itemize}

  \item \textbf{Performance is reported as a value at one operating point,
  where transfer requires a gradient.} Of the 83 full texts, 65 give
  performance at a single operating point; distance is swept in 8, source
  extent in 10, event statistics in 3, and count rate in none. A value can be
  improved indefinitely without the distance to an application changing, since
  the two are measured against different references.

  \item \textbf{The same quantity can be reported as an advantage and as a
  limitation without contradiction.} A figure given against a peer instrument
  and the same figure given against a task answer independent questions, and no
  reporting convention requires an author to state which frame is in use.
  Superiority over a predecessor and sufficiency for an application therefore
  accumulate separately, and a programme can advance genuinely in the first
  while the second does not move. No want of rigour is required for this to
  happen.

  \item \textbf{The properties that order the domains belong to the problem,
  not to the instrument.} Across five application domains (Table~\ref{tab:domains}),
  neither the size of the receiving literature, nor of the incumbent, nor of
  the stated need --- whether absolute or relative to its destination ---
  orders the domains as deployment commitment does. Two properties order them consistently:
  whether the incumbent can serve the task at all, and how many domain
  boundaries the output must cross before anyone acts on it. Both are settled
  before an instrument is designed, and neither appears in the reporting
  conventions of the field. With five domains and a commitment measure defined
  differently in each, this is an observed regularity and not a predictive
  relationship.

  \item \textbf{Where deployment commitment is concentrated is not where it is
  published.} The evidence counted here is concentrated in site
  characterisation and astrophysics; it does not establish routine service in
  either, and we found no Compton camera in routine clinical operation.
  Against declared applications of 21 works on particle therapy and 20 on
  nuclear medicine to 3 on decommissioning and field survey, \emph{the
  distribution of publication is close to the inverse of the distribution of
  deployment commitment}. The committed cases share an unknown and extended source, an
  incumbent that cannot reach the task, and a consumer who accepts a map with a
  threshold.

  \item \textbf{No stage reports the terms the next one would need.} Of 42
  works whose stage
  and bibliography could both be read, the acquisition group cites the
  reconstruction literature in 2 of 13 cases and the reconstruction group cites
  the acquisition literature in 2 of 16. Citations to the technical literature
  of verification --- the mechanism by which the receiving discipline admits a
  new measurement --- number 27 against 1616 to instruments and algorithms.
  Reviews inherit the same boundary, which is why the last stage of the chain
  --- what is delivered, to whom, within what bound --- appears in no survey
  organised by stage.

\end{itemize}

The constructive consequence is the least demanding action available and is set
out in Section~\ref{sec:metrology}: a reference object, a stated operating
point and a declared cost make results commensurable, require nobody's
permission, and carry no claim that anything previously done was wrong. The
receiving discipline took that route once already.

\section*{Data availability}
The marker definitions, the per-work screening matrix and the database queries
with their dates and counts are provided as supplementary data, so that every
figure reported here can be recomputed. The surveyed works are third-party
publications and are identified by their bibliographic record rather than
redistributed.

\section*{Acknowledgements}
This work was partially funded by ENRESA through the agreement for the
development of the research project \emph{Imagen gamma: implementaci\'on de
nuevos desarrollos e integraci\'on con dispositivos empleados por Enresa}.
J.~E. is supported by a predoctoral contract funded by Generalitat Valenciana
and European Social Fund, and by the Consejo de Seguridad Nuclear (CSN) through
the project \emph{Proton: evaluaci\'on tomogr\'afica de residuos nucleares}.

\section*{Competing interests}
The corpus surveyed here includes work by the present authors, which was
retrieved by the same queries, screened by the same criteria and scored against
the same markers as every other record, with the result reported whether or not
it is favourable. The authors declare no other competing interest.

\nocite{albiol2026bounded}
\bibliographystyle{unsrt}
\nocite{*}
\bibliography{references,references_corpus}

\end{document}